\documentclass{article}
\usepackage{spconf,amsmath,graphicx}
\usepackage{booktabs}
\usepackage{multicol, multirow}
\usepackage{tabularx}
\usepackage{amssymb}

\usepackage[stable]{footmisc}
\usepackage{url}

\usepackage[ruled,vlined]{algorithm2e}
\newcommand\tab[1][12pt]{\hspace*{#1}}

\title{ConvMixer: Feature Interactive  Convolution with Curriculum Learning for Small Footprint and Noisy Far-field Keyword Spotting}
\name{Dianwen Ng$^{1,2}$ \thanks{This work was supported by Alibaba Group through Alibaba Innovative Research Program. Award No: \textit{AN-GC-2020-010} \& Project Title: \textit{Joint approaches end-to-end system to improve far-field wake-up keyword Detection}.}, Yunqi Chen$^{1,2}$, Biao Tian$^{1}$, Qiang Fu$^{1}$, Eng Siong Chng$^{2}$}
\address{$^1$Alibaba Group, Beijing \\
$^2$School of Computer Science and Engineering, Nanyang Technological University, Singapore \\
\{dianwen.ng, jasson.cyq, tianbiao.tb, fq153277\}@alibaba-inc.com \\
aseschng@ntu.edu.sg}
\begin{document}
%
\maketitle
\begin{abstract}
Building efficient architecture in neural speech processing is paramount to success in keyword spotting deployment. However, it is very challenging for lightweight models to achieve noise robustness with concise neural operations. In a real-world application, the user environment is typically noisy and may contain reverberations. We proposed a novel feature interactive convolutional model with merely 100K parameters to tackle this under the noisy far-field condition. The interactive unit is proposed in place of the attention module that promotes the flow of information with more efficient computations. Moreover, curriculum-based multi-condition training is adopted to attain better noise robustness. Our model achieves 98.2\% top-1 accuracy on Google Speech Command V2-12 and is competitive against large transformer models under the designed noise condition.  
\end{abstract}

\section{Introduction}
\label{sec:intro}
Keyword spotting (KWS) helps to detect predetermined words in a continuous utterance. It is widely used in today’s technology to activate hands-free applications in smart devices with specific wake-up words such as \textit{``Alexa"} or \textit{``Hey Siri"}. In most cases, these gadgets are constraint with low memory and computational resources. Hence, it is important to consider the feasibility of the system with the emphasis on low computational cost and with reasonable model size. Recent works on the small footprint KWS model \cite{majumdar2020matchboxnet, rybakov2020streaming} have gained massive successes on less noisy and close-talking audio sets. However, the system becomes vulnerable, particularly in the scenario of far-field speech with a low signal-to-noise ratio (SNR). It is evident that small models \cite{arik2017convolutional, prabhavalkar2015automatic} with lower networks complexity face a tougher challenge in generalizing noisy signals. As a result, the accuracy of the system is likely to deteriorate causing bad user experience when devices get less responsive or subjected to a higher false alarm rate. 

Prior works on improving the overall performance and noise robustness include using an attention-based module to boost the efficiency of the audio networks \cite{rybakov2020streaming, zhang2016end, jung2020multi}. This provides the ability to selectively focus on valuable segments of the audio sequence. Furthermore, self-attention such as the audio transformer has shown to outperform the convolutional networks-attention hybrid \cite{gong2021ast, berg2021keyword}. Nevertheless, the huge computational and memory complexity eminently discounts its usability on small devices and becomes less desirable .

In this paper, we focus on the actual application scenario of a noisy far-field environment. We attempt to optimize the performance of a small KWS system by constructing a novel convolutional networks (CNN) encoder with a mixer module that offers a strong alternative to attention. The mixer unit computes the weighted feature interaction of the global channel to allow the flow of information with varying importance. Prominently, the CNN encoder has a light memory footprint and is highly effective at smaller model sizes. Furthermore, we proposed a learning strategy with curriculum-based multi-condition training that surpasses the vanilla multi-condition learning to achieve better noise robustness. We have shown from our experiments that our system outperforms the existing state-of-the-art (SOTA) solutions for small footprint KWS under the noisy far-field condition. Besides, the performance of our proposed system is comparable to models of its size 50 times larger. 


\section{Related works}
\label{sec:prior}
\textbf{Small Footprint Keyword Spotting} - 
Deep neural networks (DNN) has been proven to be effective in KWS task \cite{chen2014small}. With the rapid development of CNN to automatically learns the encoding of spatial information given a sequence, it has become increasingly popular in acoustic modelling. Earlier work \cite{sainath2015convolutional} has demonstrated the use of CNN to execute small footprint KWS. Subsequently, \cite{chollet2017xception, zhang2017hello} have extensively reduced the memory footprint with depthwise separable convolution and achieved the best model size accuracy tradeoff. 

\noindent \textbf{Noise Robust Speech Model} - 
Multi-condition training has emerged as the method of choice for its simple strategy for noise robustness in small footprint model. 
However, it gets incompetent when the model learns from a broader range of noises, i.e. from a very low SNR such as -10 dB to clean \cite{CL2}. Recently, \cite{CL2, CL1} have proposed a more effective method with curriculum learning. In short, they train the model starting with clean or high SNR audio and then gradually increases the noise level to lower SNR. This progressive training is more effective than the conventional method in obtaining noise robustness.


\section{Methodology}
\label{sec:method}
\subsection{Model Architecture}
\label{ssec:model}
\begin{figure}[t]
    \centering
    \includegraphics[width=0.63\textwidth, angle =270]{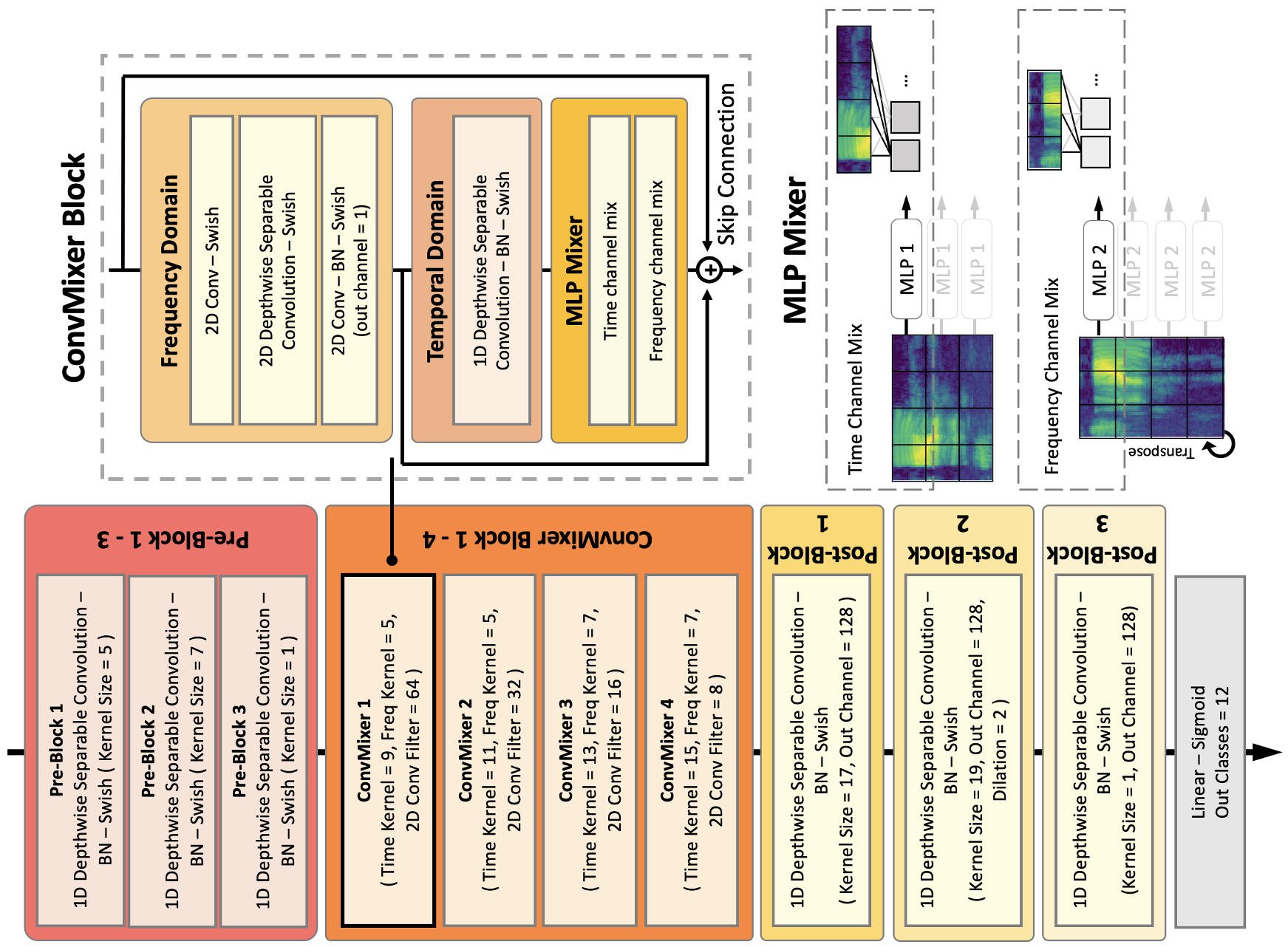}
    \caption{Overview of our ConvMixer model architecture}
    \label{fig:model}
\end{figure}

Our ConvMixer networks consist of three main sections, i.e. pre-convolutional block, convolution-mixer block and post-convolutional block. Similar to the previous work, we built our model encoder based on depthwise separable (DWS) convolution as it provides the most efficient computation using a small number of model parameters. We designed our pre and post convolutional blocks with the same neural layers of a 1-dimensional DWS, batch normalization followed by the swish activation \cite{ramachandran2017searching}. All of the following blocks are convolved with different kernel sizes as listed in Fig. \ref{fig:model} and padded to preserve the dimension from the previous time frame. However, \cite{kim2021broadcasted} discussed that the property of translation equivariance for the convolutional operation in 1D is not preserved in the frequency domain. This would compromise the learning of some spatial information along with the frequency channel. Hence, we consider introducing 2-dimensional DWS, specifically in our ConvMixer block. 

The ConvMixer block takes the previous channel $\times$ time feature and passes it through the 2D convolutional sub-block for frequency domain extraction. This creates a third dimension that expresses the rich information from the frequency domain. To maintain the shape from the previous input, we employed a pointwise convolution that compresses it back to fit the shape. Then, we implemented the temporal domain feature extraction with a 1-dimensional DWS block. The product from these two operations will result in frequency and temporal rich embeddings. Following that, we built a mixer layer to allow the flow of information over the global feature channel. Lastly, we added skip connections from the previous output and the 2D feature connecting to the output of the block. We express our ConvMixer block in the following equations: \begin{equation}
\begin{split}
\label{eq:1}
z &= \sigma \circ \boldsymbol{f_1}(\sigma \circ f(x)) \\ 
y_1 &= \sigma \circ \text{BatchNorm} (f(z))
\end{split}
\end{equation}
\begin{equation}
\label{eq:2}
y_2 = \sigma \circ \text{BatchNorm}( \boldsymbol{f_2} (y_1))
\end{equation}
\begin{equation}
\label{eq:3}
\tilde{y} = x + y_1 + \boldsymbol{f_3}(y_2)
\end{equation} where eq (\ref{eq:1}) computes the frequency domain features with $\boldsymbol{f_1}$ as the 2d-DWS, 2D convolution function $f$. Eq (\ref{eq:2}) computes the temporal domain features with $\boldsymbol{f_2}$ as the 1d-DWS. Eq (\ref{eq:3}) computes the output of the block with $\boldsymbol{f_3}$ as the mixer layer and $\sigma$ as the swish activation for eq (1-3).

\subsection{Mixer Layer}
\label{ssec:convmixer}
The attention layer is trendy for its strength to allow networks to focus on useful spatial information. Nonetheless, this requires heavy linear computation. Instead of weighing the relevance of an element to every other token, \cite{lee2021fnet, tolstikhin2021mlp} suggested mixing the token channel-wise as an alternative approach to feature communication. Therefore, we proposed to utilize two types of multi-layer perceptrons (MLP), namely temporal channel mixing and frequency channel mixing, to induce the interaction between the feature space. Each MLP mixing involves two linear layers and a GELU activation unit independent of each temporal and frequency channel. This is defined as \begin{equation}
\begin{split}
\label{eq:4}
u_{*, i} &= x_{*, i} + W_2 \cdot \delta(W_1 \cdot \text{LayerNorm}(x)_{*, i}) \\
y_{j, *} &= u_{j, *} + W_4 \cdot \delta(W_3 \cdot \text{LayerNorm}(u)_{j, *})
\end{split}
\end{equation} where $\delta$ represents the GELU unit. $W_1$ and $W_2$ are the learnable weights of the linear layers for temporal channel shared across all frequency $i$, for $i \in \{1, I\}$. $W_3$ and $W_4$ are the learnable weights of the linear layers for frequency channel shared across all $j$, for $j \in \{1, J\}$.

As illustrated in Fig. \ref{fig:model}, we only learn the weights that connect the channel feature with the weighted coefficient shared across the other domain. For convenience, we transpose the latent feature for frequency channel mix so that the arithmetic stays the same as temporal channel mix. Following that, another transpose will be done to recover its original frequency $\times$ time arrangement. The learned coefficient value facilitate the distribution of information with different significance similar to the attention but to be much more computationally efficient. 

\subsection{Curriculum Based Multi-condition Training}
\label{ssec:curriculum}

To enhance the noise robustness of our model, the aforementioned curriculum learning based on the SNR level is employed as a training strategy.
To execute, we divide the training process into five progressively harder steps. At the start, we conditioned the model on clean samples without noise. In the following three steps, noises will be introduced to the fixed $N$ samples in increments of -5dB, and all the conditions in $N$ samples is uniformly distributed, i.e. [clean, 0], [clean, 0, -5], [clean, 0, -5, -10]. Lastly, we include far-field audio by augmenting half of our dataset with room impulse response (RIR) data.

In every epoch of each stage, we record the learning progress with the validation accuracy and the loss. Next, the progression step criterion $c$ is defined as the difference between the normalized validation accuracy and loss. Normalization is based on the accuracy and loss of previous epochs. Eq (\ref{eq:5}) depicts the general arithmetic for computing the $m^{th}$ epoch value of the normalized accuracy and loss. Note that the normalization result is zero if $m$ is equal to zero. Subsequently, if $c$ is not higher than the current best criterion for a consecutive of 10 epochs, the model with the latest best criterion will be loaded and progressed to the next stage of difficulty for training. The 
complete training strategy is shown in Algorithm \ref{algo}. \begin{equation}
\label{eq:5}
Norm(a_m)=\frac{a_m-min(A)}{max(A)-min(A)}, A=\{a_1,a_2...,a_m\} 
\end{equation}

\begin{algorithm}[ht]
\KwIn{clean audio utterances, $D = \{x_i, y_i\}_{i=1}^N$}
\nl\textbf{Initialize:} $bst\_crit = 0$; stage $= 0$;\\
\tab\tab\tab model parameters, $F(\Theta)$; \\
\nl \textbf{while} $\text{stage} < 5$ \textbf{do} \\
    \nl \tab \textbf{for} epoch, $m = 1, 2, \ldots$ M \textbf{do}\\
    \nl \tab \tab $\hat{y_m} = \text{Forward}(F_m(x, \Theta))$; \\
    \nl \tab \tab $loss_m = \text{BCE}( \hat{y_m}, y)$;  \\
    \nl \tab \tab $acc_m = \text{Accuracy Score}(\hat{y}, y)$; \\
    \nl \tab \tab \textbf{compute} $c = Norm(acc_m)-Norm(loss_m)$; \\
    \nl \tab \tab \textbf{update} $bst\_crit \leftarrow max(bst\_crit, c)$ \\
    \nl \tab \tab Saving best model if c == \textit{bst\_crit}; \\
    \nl \tab \tab \textbf{if} $c < bst\_crit$ for 10 epochs \textbf{then} \\
    \nl \tab \tab \tab \textbf{update} stage $\leftarrow$ stage $+ 1$;\\
    \nl \tab \tab \tab Load best model from previous stage;\\
    \nl \tab \tab \tab Augment noise with next level of difficulty; \\
\caption{\bf Curriculum Based Multi-condition} 
\label{algo}
\end{algorithm}

\section{Experiments}
\label{sec:experiments}
\begin{table*}[!htbp] \footnotesize \centering \setlength\tabcolsep{8.8pt} 
\begin{tabular}{@{}l|c|c|c|ccccc@{}}
\toprule
\multicolumn{1}{c|}{\multirow{2}{*}{Model}} &
  \multirow{2}{*}{\begin{tabular}[c]{@{}c@{}}Num. of \\ Params (K)\end{tabular}} &
  \multirow{2}{*}{MACs (M)} &
  \multirow{2}{*}{\begin{tabular}[c]{@{}c@{}}Acc. of V2-12\\ Official (\%)\end{tabular}} &
  \multicolumn{5}{c}{Accuracy of Far-field Test Command, SNR in dB (\%)} \\  \cline{5-9}
\multicolumn{1}{c|}{} &        &        &       & Clean & 20 dB    & 0 dB     & -5 dB    & -10 dB   \\ \midrule
MHAtt-RNN \cite{rybakov2020streaming}             & 784   & 141.6 & 98.04 & 78.83 & 74.25 & 61.98 & 55.87 & 50.98 \\
KWT-1 \cite{berg2021keyword}                  & 607   & 53.8  & 97.72 & 88.73 & 85.46 & 73.52 & 67.59 & 59.07 \\
ResNet-15 \cite{tang2018deep}             & 238   & 961.2    & 96.48 & 89.45 & 87.34 & 79.00 & 73.58 & 66.73 \\
MatchboxNet-6x2x64 \cite{majumdar2020matchboxnet}     & 140   & 36.8  & 97.60 & 87.34 & 85.19 & 75.58 & 70.06 & 62.35 \\
ConvMixer (Ours)       & 119   & 22.2  & 98.20 & 90.38 & 87.85 & 78.10 & 72.78 & 66.50 \\
ConvMixer $\dagger$ (Ours)  & 119   & 22.2  & \textbf{98.20} & \textbf{93.16} & \textbf{90.83} & \textbf{83.04} & \textbf{78.39} & \textbf{71.88} \\ \midrule
AST-Tiny \cite{gong2021ast}               & 5,805 & 782.2 & 97.65 & 91.02 & 87.71 & 83.31 & 78.95 & 72.32 \\
KWT-3 \cite{berg2021keyword}                 & 5,361 & 526.3 & 98.54 & 93.47 & 91.08 & 83.97 & 78.45 & 71.08 \\
\bottomrule
\end{tabular}
\caption[Comparisons with the SOTA model.$\dagger$: Proposed model with curriculum learning. MACs computed with link]{Comparison with the SOTA models (\textbf{$\dagger$}: proposed model with curriculum learning). MACs computed with \footnotemark.}
\label{main-exp}
\end{table*}

\subsection{Experimental Setup}
\label{ssec:setup}

\subsubsection{Dataset for Far-field Keyword Spotting}
\label{sssec: dataset}
We evaluate our proposed system on the Google Speech Commands V2 \cite{speechcommandsv2}. It contains 105,000 utterances of 35 unique words, each of 1 second long, sampled at 16 kHz. We use the official train, validation and test split provided for the 12 labels classification task. This covers the words: `up', `down', `left', `right', `yes', `no', `on', `off', `go' and `stop' together with `silence' and `unknown' classes. The latter class is treated from the remaining words in the dataset.

To simulate our noisy far-field environment, we have employed two additional datasets. We apply the noise samples from MUSAN \cite{musan}, where it contains 930 files of assorted noises sampled at 16kHz, with a total duration of about 6 hours. These carry various
technical and non-technical noises such as DTMF tones, thunder and car horns and we add them to our commands to mimic the audio under different noisy conditions. 
Far-field speech is generated using the reverberation from BUT Speech@FIT Reverberation Database \cite{RIRdata}. The dataset holds the RIR data from nine rooms of different sizes (large, middle and small sizes). 

\subsubsection{Implementation Details}
\label{sssec: preprocessing}
\textbf{Input Feature} - We use the input features of a 64-dimensional log Mel filterbank (FBank) with a 25ms window size and a 10ms shift. We fixed the resolution of our FBank at 98 $\times$ 64, equivalent to 1s of the utterance. Commands that are shorter than 1s will be zero-padded to the right. During training, data augmentation is performed with a time shift in the range of -100 to 100ms. Furthermore, spectrogram masking with both the time and frequency masking parameters of max length 25 is adopted. We generate our noisy data with SNR chosen from the list of set [0, -5, -10] dB as detailed in section \ref{ssec:curriculum}. Then, for stronger learning regularization, input mixup is executed with a mixup ratio of 0.5 on the training samples.

\noindent
\textbf{Model Training} - Model is trained with a batch size of 128 and an initial learning rate of 6e-3 factored by 0.85 on every four epoch intervals after the fifth epoch. Adam optimizer and binary cross-entropy loss are used in the optimization process. We trained our model for 200 epochs with early stopping criteria defined as in the progression step criterion in section \ref{ssec:curriculum}. 

\subsection{Results}
\label{ssec:results}
We compare the performance of the ConvMixer with previously proposed SOTA models. Models are retrained from the official source code provided with our designed data environment. The results are shown in Table \ref{main-exp}. 
From the table, we observed that our proposed model achieved the SOTA accuracy among small models when tested on the official V2-12. Furthermore, it has a noticeable drop in the number of model parameters and MACs that signify lower memory and computation resources. 
Most importantly, when evaluated on the noisy far-field condition, we scored an absolute improvement of 3\% against MatchboxNet with a similar memory footprint of the same multi-condition training. This is extended to 7.4\% for our curriculum-based training. Finally, we show that the proposed model is competitive against the larger transformer-based model (KWT-3, AST-Tiny) under the challenging noisy far-field conditions.


\footnotetext{\scriptsize \url{https://github.com/sovrasov/flops-counter.pytorch}}



\subsection{Ablation Studies}
\label{ssec:ablation}
We further investigate the importance of the feature interactive structure: MLP mixer under noisy far-field conditions. Using the same curriculum based multi-condition training method, we removed the MLP mixer in the ConvMixer block of our model and obtained the results as shown in Table \ref{mixer_comp}.The addition of the Mixer layers provide a substantial boost in the accuracy of approximately 7\%, indicating the usefulness of this feature interactive structure in making the model more robust.
\begin{table}[!htbp] \centering \footnotesize \setlength\tabcolsep{4.0pt} 
\label{compare mixer}
\begin{tabular}{@{}lccccc@{}}
\toprule
\multirow{2}{*}{ConvMixer $\dagger$} & \multicolumn{5}{c}{\begin{tabular}[c]{@{}c@{}}Accuracy of Far-field Test Command (\%)
\end{tabular}} \\ \cmidrule(l){2-6} 
                  & Clean  & 20dB   & 0dB    & -5dB   & -10dB  \\ \midrule
With MLP Mixer    & \textbf{93.16} & \textbf{90.83} & \textbf{83.04} & \textbf{78.39} & \textbf{71.88}     \\
Without MLP Mixer & 85.77 & 83.52 & 76.56 & 72.60 & 66.26 \\ \bottomrule
\end{tabular}
\caption{Comparison with/without MLP mixer layer}
\label{mixer_comp}
\end{table}

\begin{figure}[!h]
    \centering
    \includegraphics[width=0.48\textwidth]{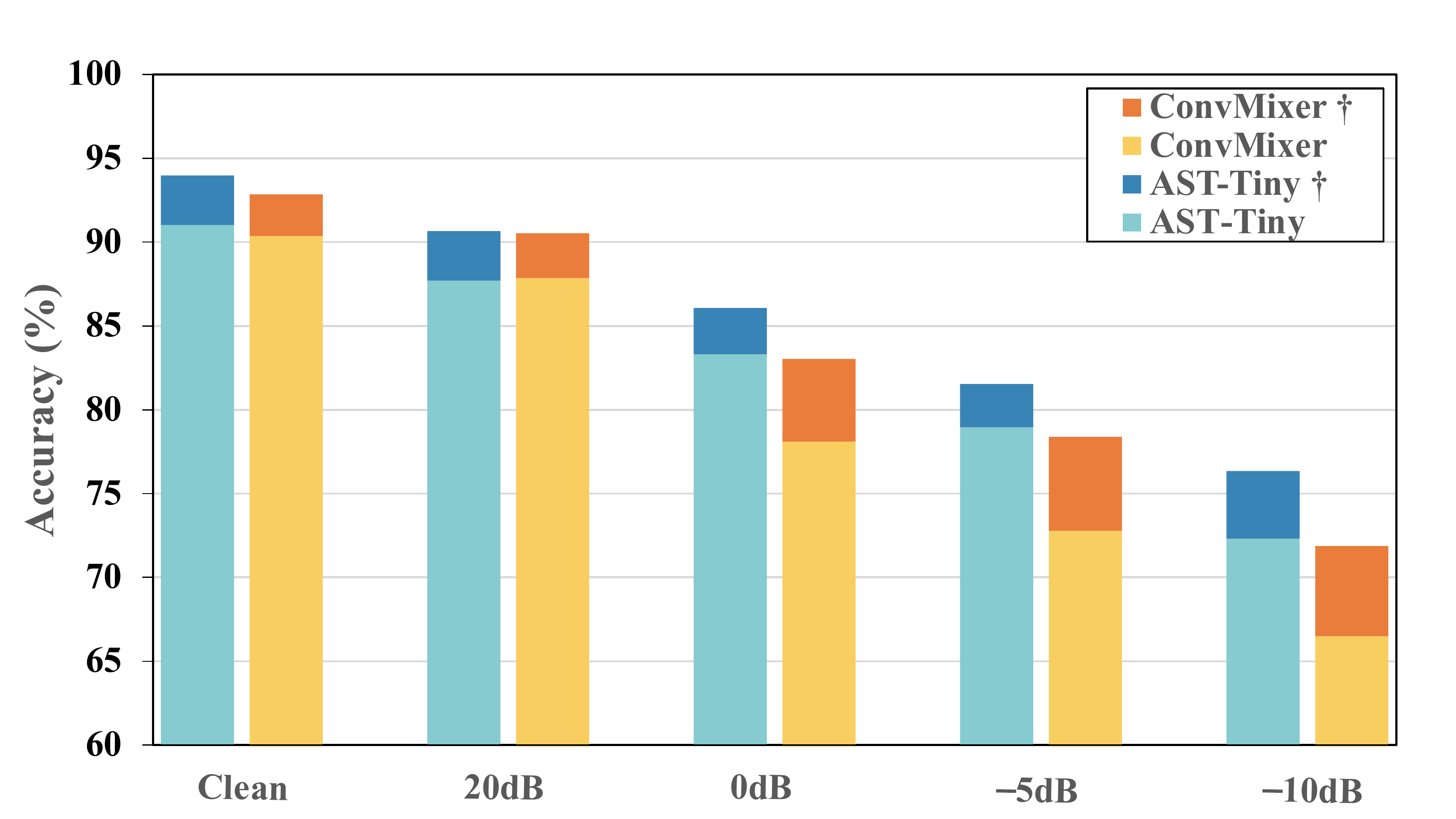}
    \caption{Performance gains from curriculum learning}
    \label{fig:CL_Compare}
\end{figure}

We also explored the performance gains from curriculum based multi-condition training on the transformer based AST-Tiny and the results are shown in Fig \ref{fig:CL_Compare}. Curriculum learning on \textbf{AST-Tiny $\dagger$} leads the chart with an improvement in accuracy of about 3\% compared to multi-condition training. This is in agreement with the capability of curriculum learning to improve the performance of the model. 
Despite that, our proposed model only lags less than 2\% behind \textbf{AST-Tiny $\dagger$}. 
Also, the chart shows that curriculum learning is more effective on \textbf{ConvMixer $\dagger$} with smaller model parameters, especially in lower SNRs, boosting accuracy by about 5.5\%. 

\vspace{-0.42cm}  
\vspace{-0.8cm}

\section{Conclusion}
\label{sec:conclusion}
In this work, we introduce a novel small footprint model \textit{ConvMixer} with the feature interactive structure \textit{MLP mixer}. 
Curriculum based multi-condition training method is applied to improve noise robustness. 
The performance of our ConvMixer exceeds the existing SOTA KWS in clean and noisy far-field conditions on Command V2-12. Furthermore, it also matches the performance of the transformer-based KWS that uses 50 times more memory consumption and computing resources.
The results highlight the potential of ConvMixer used in deployment at the endpoint and application in real-world scenarios.





\bibliography{refs}

\end{document}